\documentstyle[12pt]{article}

\newcommand{\beq}{\begin{equation}}
\newcommand{\eeq}{\end{equation}}
\newcommand{\beqcol}{\begin{array}{rcl}}
\newcommand{\eeqcol}{\end{array}}

\begin{document}
{\thispagestyle{empty}
\begin{flushright}  YITP/U-95-12  \end{flushright}
\vfill
\begin{center}
{\Large 
%{\obeylines
Selfdual 2-form formulation of gravity and \\
classification of energy-momentum tensors  
}
\\
\vspace{2cm}
{\sc Mathias Pillin}\footnote{JSPS-fellow, supported also by Alexander
von Humboldt-foundation;\newline e-mail: map@yisun1.yukawa.kyoto-u.ac.jp}  
\\
\vspace{1cm}
Uji Research Center             \\
Yukawa Institute for Theoretical Physics  \\ 
Kyoto University                          \\
Uji, Kyoto 611, JAPAN     \\ 

\medskip

\vfill
{\bf Abstract}
\end{center}
\begin{quote}
It is shown how the different irreducibility classes of the 
energy-momentum tensor allow for a Lagrangian formulation 
of the gravity-matter system using a selfdual 2-form as 
a basic variable. It is pointed out what kind of 
difficulties arise when attempting to construct 
a pure spin-connection formulation of the gravity-matter 
system. Ambiguities in the 
formulation especially concerning the need for constraints are clarified.  
\end{quote}

PACS: 0420

\eject
}

\setcounter{page}{1}

\section{Introduction}

\bigskip

Recently there has been interest in formulations of gravity 
using self-dual two forms as basic variable
\cite{CDJ}. This approach allows for an almost complete elimination of 
the metric in the action, leaving the self-dual part of 
the spin connection and a scalar density as the only 
gravitational quantities in the theory. This formulation 
provides a natural covariantization of Ashtekar's 
canonical formalism \cite{ASHT}. 

We briefly review the procedure for vacuum general relativity. The 
action as function of the basic two-form $\Sigma^{AB}$ \cite{PLEB}, 
the Weyl part of the curvature $\Psi_{ABCD} = \Psi_{(ABCD)}$, 
and the spin-connection  one-form $\omega_{AB}$ with 
$ R_{AB} = {\rm d} \omega_{AB} + \omega_{AC} \wedge 
\omega^{C}_{\ \ B} $ is given as follows:

\beq
S[\Sigma^{AB},\omega_{AB},\Psi_{ABCD}] = 
  \int \Sigma^{AB} \wedge R_{AB} 
  - {1\over 2} \Psi_{ABCD} \Sigma^{AB} \wedge \Sigma^{CD}
\label{action1}
\eeq

Note that summation convention is used in this paper. Capital latin 
letters range over 0,1. They are raised and lowered according to 
the rules in \cite{PR}. Symmetrization and anti-symmetrization in 
indices are denoted by $(AB)$ and by $[AB]$ respectively. 
The indices coming from the second fundamental representation of 
$SL(2,{\bf C})$ are denoted by $\dot{A}, \dot{B}$ etc. However, in 
the present formulation care has to be taken of the 
reality conditions \cite{CDJ}.
\medskip

The variation of the action with respect to $\Psi_{ABCD}$, 
$\omega_{AB}$, and $\Sigma^{AB}$ yields the equations of motion:

\beq
\Sigma^{(AB} \wedge \Sigma^{CD)} = 0, \quad
{\rm D}\Sigma^{AB} = 0, \quad
R_{AB} = \Psi_{ABCD} \Sigma^{CD}.
\label{eqmot}
\eeq

D denotes the covariant derivative with respect to $\omega_{AB}$. 
The first equation shows that the self dual two-form can be 
expressed in terms of basic tetrads, i.e. 
$\Sigma^{AB} = \theta^{A}_{\ \dot{A}} \wedge  \theta^{B\dot{A}}$. 
The fact that the covariant derivative applied to 
$\Sigma^{AB}$ vanishes shows that $\omega_{AB}$ is the 
self-dual part of the spin-connection. The last equation in 
(\ref{eqmot}) is just the Einstein equation in the vacuum.  

The metric independent formulation can now be obtained by 
first solving (\ref{eqmot}) for $\Sigma^{AB}$ and then 
eliminating $\Psi_{ABCD}$ under the condition that the 
it has a vanishing trace. The latter is only possible 
if $\Psi_{ABCD}$ is invertible. Thus the present 
formulation does not allow vacuum spacetimes of Petrov type 
$\{ 31\}$, $\{ 4\}$, and $\{ -\}$. Hence it is not clear if this procedure is 
reasonable when attempting to quantize the theory since under 
certain circumstances the Petrov type might change, see 
e.g. \cite{PetCh}. 

The final form of the action \cite{CDJ} then involves only $\omega_{AB}$ 
and a scalar density $\eta$ of weight --1. 
\beq
S[\omega_{AB}, \eta ] = \int \eta (\epsilon R^{AC} \wedge R^{BD} ) 
 (\epsilon R_{AB} \wedge R_{CD} )
\label{action2}
\eeq

In this paper a systematic treatment of matter couplings 
to gravity using the self-dual 2-form $\Sigma^{AB}$ as 
basic variable will be presented. It will be shown how 
the different irreducibility classes of the energy-momentum 
tensor of matter enter in a Lagrangian formulation of the 
problem. Therefore this treatment is general since it 
does not rely on some specific examples of matter 
couplings.

The outline of the paper is as follows. In section 2 we recall 
some basic facts about Einstein theory in spinorial form and 
point out the problems and objectives in a Lagrangian formulation 
using $\Sigma^{AB}$ as basic variable. The different irreducibility 
classes of the energy-momentum tensor are presented in section 3. 
Since we want to treat $\Sigma^{AB}$ as a form the matter 
degrees of freedom have to be embedded in differential forms. 
This embedding introduces undesired degrees of freedom 
which have to be projected out. The constraints causing 
this projection are calculated in section 4. In section 5 the 
the matter Lagrangians are formulated. Finally a few remarks on 
the physically important cases of scalar- and spinorial- 
matter actions are made in section 6.

\bigskip
\bigskip

\section{Matter Couplings}

\bigskip

In this section facts about the spinorial formulation \cite{PR} of 
Einstein theory are recalled since they will be needed in the 
remainder of the paper. 

\medskip

The curvature 2-form $R_{AB}$ which is the (anti-) self-dual 
part of the Riemann curvature can be regarded as a Lorentz 
or $SL(2,{\bf C})$ tensor respectively. Denoting 
curved space-time indices by greek letters we can 
give the decomposition of this tensor into $SL(2,{\bf C})$ 
irreducible parts. Denoting by $\varepsilon$ the $SL(2,{\bf C})$ 
invariant skew symbol we get (cf. \cite{PR}):

\beq
R_{AB\ \gamma \delta} = R_{AB\ C\dot{C} \ D \dot{D} } =  
\Psi_{(ABCD)} \varepsilon_{\dot{C}\dot{D}} + 
\Phi_{(AB)(\dot{C}\dot{D})}  \varepsilon_{CD} 
+ \Lambda (\varepsilon_{AC}\varepsilon_{BD}  +  
     \varepsilon_{AD}\varepsilon_{BC}) \varepsilon_{\dot{C}\dot{D}} 
\label{riemann}
\eeq
Here $\Psi_{(ABCD)}$ corresponds as mentioned above to 
the Weyl curvature, $\Phi_{(AB)(\dot{C}\dot{D})}$ is the 
traceless Ricci tensor, and $\Lambda$ contains 
the scalar parts of the Riemann tensor. 
Making use of the decomposition of the product 
of the tetrad 1-forms cf. \cite{PLEB}

\beq
\theta^{A\dot{A}} \wedge \theta^{B\dot{B}} = 
\epsilon^{AB} \tilde{\Sigma}^{\dot{A}\dot{B}} + 
\epsilon^{\dot{A}\dot{B}} \Sigma^{AB}, 
\label{expan1}
\eeq 

we get:

\beq 
R_{AB} = \Psi_{(ABCD)} \Sigma^{CD} 
       + \Phi_{(AB)(\dot{C}\dot{D})}\tilde{\Sigma}^{\dot{C}\dot{D}}   
       + 2 \Lambda \Sigma_{AB}.
\label{riemann-form}
\eeq

For later purposes we rewrite the Einstein equations in 
the $SL(2,{\bf C})$ version. The matter part entering 
these equations is described by the energy-momentum tensor 
$T_{\alpha\beta}= T_{\beta\alpha}$. Due to its symmetries 
this tensor admits the following decomposition:

\beq 
T_{\alpha\beta} = T_{A\dot{A} B\dot{B}} = 
 T_{(AB) (\dot{A}\dot{B})} 
   + \epsilon_{AB}\epsilon_{\dot{A}\dot{B}}{1\over{4}} 
           T_{\alpha}^{\alpha} 
\label{tdecomp}
\eeq

Obviously the first summand is the traceless part whereas 
the second one corresponds to the scalar part of $T_{\alpha\beta}$. 

The Einstein equations with cosmological constant contribution 
$\lambda$ and gravitational coupling $\gamma$ are 
$R_{\alpha\beta} -{1\over{2}} R g_{\alpha\beta} + 
 \lambda g_{\alpha\beta} = -8 \pi \gamma T_{\alpha\beta} $. 
Completely equivalent equations are obtained using the 
$SL(2,\bf{C})$ decomposition by a straightforward calculation: 

\beq
\Phi_{(AB)(\dot{C}\dot{D})} = 4\pi \gamma T_{(AB) (\dot{A}\dot{B})}, 
\qquad 
\Lambda = {1\over{6}} \lambda + {1\over{3}} \pi \gamma
T_{\alpha}^{\alpha}
\label{einstein}
\eeq

To be complete we mention that the Bianchi identities 
establish differential relations among 
the irreducible parts of the the Riemann tensor. Denoting 
by $\nabla_{A\dot{A}}$ the covariant derivative these 
relations are simply $\nabla_{\dot{B}}^A \Psi_{(ABCD)} = 
\nabla^{\dot{A}}_{(B} \Phi_{(CD))(\dot{A}\dot{B})}$ and 
$\nabla^{A \dot{A}}\Phi_{(AB)(\dot{A}\dot{B})} + 
 3  \nabla_{B \dot{B}}\Lambda  = 0 $.

\bigskip

The problem to be considered in this paper is how to 
couple matter to gravity in an action formulation 
which uses the 2-form $\Sigma^{AB}$ as a basic variable 
not only in the gravity part (as outlined in the 
introduction) but also in the matter part. 

We will consider actions of the form $S_{total} = 
S_{gravity} + S_{matter}$. In the remainder of 
the paper we do not mention the exterior product 
explicitly. Up to constants this action 
takes the form with $\phi$ denoting matter degrees 
of freedom:

\beq
S_{total} = \int \Sigma^{AB} R_{AB} - {1\over{2}} \Psi_{(ABCD)} 
\Sigma^{AB}\Sigma^{CD} -{1\over{6}} \lambda \Sigma^{AB}\Sigma_{AB} 
+ S_{matter}[\Sigma^{AB},\phi]
\label{genaction1}
\eeq

$S_{matter}$ has to be formulated in such a way that the 
variation by $\Sigma^{AB}$ and maybe having used the matter 
equations of motion the Einstein equations are obtained 
in the form (\ref{einstein}) after expanding $R_{AB}$ 
(\ref{riemann-form}). For the case of the Yang-Mills-gravity system this 
has been done in \cite{ROB}. A detailed treatment can be found 
in section 5. 

\medskip

In general if a cosmological constant is present or 
matter is coupled to gravity the self-dual 
formulation of section 1 and the elimination of the metric degrees of 
freedom become difficult \cite{CDJ}, see also \cite{PEL,ASHT1}. 

\medskip

The origin of the problem is due to the need for 
a constraint that ensures the tracelessness of the Weyl tensor 
in the presence of a cosmological constant or of matter 
fields. We review briefly the first case (see the last 
ref. in \cite{CDJ} and \cite{PEL}) while the problems 
in the matter coupled case are addressed at the end 
of section 5. For the moment we adopt a kind of 
index free notation.

One can rewrite the action (\ref{genaction1}) without matter 
contributions using $X:= \Psi + \lambda /3$ and a Lagrange 
multiplier $\mu$:

\beq
S = \int \Sigma R - {1\over{2}} X \Sigma\Sigma 
    + \mu ({\rm tr}X -\lambda )
\label{revact1}
\eeq
The last term is needed to ensure the tracelessness of 
the Weyl tensor. 

Applying an identity \cite{CDJ} which holds for any 3x3-matrix 
to $X$ and solving the $\Sigma$ equations of motion we arrive 
at an action of the form:

\beq 
S = \int {1\over{2}} {\rm{tr}}(X^{-1}M) 
   + \rho ( ( {\rm{tr}}X^{-1} )^2 - ({\rm{tr}}X^{-2}) 
            - 2 ({\rm{det}}X)^{-1} \lambda ) 
\label{revact2}
\eeq
We define \cite{CDJ} $M_{ABCD}:= R_{AB}\wedge R_{CD}$ and 
$\rho = \mu {\rm{det}}X /2$. To obtain a pure spin-connection 
formulation one has to eliminate X which can be done by 
considering the variation of (\ref{revact2}) by $X^{-1}$. This 
yields:

\beq
M = 4\rho ( X^{-1} - {\rm{tr}}X^{-1} {\bf 1} 
      + \lambda ({\rm{det}}X)^{-1} X )
\label{revact4}
\eeq
It has been pointed out in \cite{CDJ} that (\ref{revact4}) is 
difficult to solve for $X^{-1}=X^{-1}(M,\rho)$ but in 
\cite{PEL} and in the last ref. of \cite{CDJ} a 
partially satisfactory solution to this problem is 
proposed. One assumes the existence of a solution 
$Y(M,\rho) = X^{-1}(M,\rho)$. Inserting (\ref{revact4}) 
with $Y$ into (\ref{revact2}) one can show that 
the resulting action is a functional of ${\rm{tr}}M$, 
${\rm{tr}}(M^2)$, ${\rm det}M$, and $\rho$. These constituents 
can be related to $Y$. In the course one has to solve a 
quadratic equation which leads to the following 
actions, $\chi:= \lambda/(8\rho)$:

\beq
S = {1\over{2\lambda}} \int \chi^{-1} (  ( 1+ \chi {\rm tr} M) 
 \pm( (1+ \chi {\rm tr} M )^2 - 2\chi^2 ( {\rm tr}M^2 -{1\over 2}
 ({\rm tr}M)^2 ) + 8 \chi^3 ({\rm{det}}M))^{1\over 2} )
\label{revact5}
\eeq

However, the so obtained action is not unique.  

\bigskip

For certain types of matter couplings and in the 
presence of a cosmological 
constant the action (\ref{action1}) has been modified in \cite{CDJ} 
to account for the additional physical information: 
\beq
\int \Sigma^{AB} \wedge \Gamma_{AB} 
    - {1 \over 2} \Xi_{ABCD}  \Sigma^{AB} \Sigma^{CD} 
    + {1 \over 2} ( \Xi_{AB}^{\ \ \ \ AB} - \Delta  )\Sigma_{CD}\Sigma^{CD}
\label{action3}
\eeq
In this expression $ \Gamma_{AB} = R_{AB} + M_{AB} $ with 
$M_{AB} $ describing one part of the matter action. $\Xi_{ABCD}$ 
contains the Weyl part of the curvature and another part 
of the matter action. This object may not be traceless and 
therefore causes problems (see {\it Erratum} in \cite{CDJ}). 
$\Delta$ denotes additional trace parts coming from the 
matter action or the cosmological constant. In section 5 
we will present a general framework for an action comparable to 
(\ref{genaction1}) and (\ref{action3}) corresponding to the 
irreducibility classes of the energy-momentum tensor of matter.

\bigskip
\bigskip

\section{Classification of the energy-momentum tensor}

\bigskip

The coupling of matter to gravity leads to Einstein 
equations with a nonvanishing energy-momentum tensor. It is 
possible to give an algebraic classification \cite{PR,LS} of both the 
energy-momentum tensor and the Ricci tensor. The latter 
contains direct information of the matter part by 
the field equations (\ref{einstein}).

We consider the traceless part of the hermitian energy-momentum
tensor: $T_{AB\dot{C}\dot{D}}=T_{(AB)(\dot{C}\dot{D})}$. The 
first step in the classification is the reducibility 
of this object. 

\bigskip

\begin{tabular}{lllclc}
    &               &            &    & & Real dimension \\[0.8ex]
A.    &   $  (2,2) $  & $T_{AB\dot{C}\dot{D}}$  & \  & irreducible & 9 \\[0.8ex]
B1.   &   $ (1,1)(1,1)  $  & 
      $T_{AB}^{\dot{C}\dot{D}}$ &=& $\Gamma^{(\dot{C}}_{(A}
                                 \Lambda^{\dot{D})}_{B)}   $  & 7  \\[0.8ex]
B2.   &   $ |(1,1)|^2  $  & 
      $T_{AB}^{\dot{C}\dot{D}}$ &=& $\Gamma^{(\dot{C}}_{(A}
                                 \overline{\Gamma}^{\dot{D})}_{B)}$ & 7  \\[0.8ex]
C.    &   $ (1,1)^2  $  & 
         $  T_{AB}^{\dot{C}\dot{D}}$ & =& $\pm \Lambda^{(\dot{C}}_{(A}
                                 \Lambda^{\dot{D})}_{B)}    $ & 4  \\[0.8ex]
D.    &   $ (1,1) |(1,0)|^2  $ & 
             $T_{AB\dot{C}\dot{D}}$ &=& $\rho_{(A}\Lambda_{B)(\dot{C}}
                                \bar{\rho}_{\dot{D})}     $ &  6 \\ [0.8ex]   
E.    &   $ |(1,0)(1,0)|^2 $    & $T_{AB\dot{C}\dot{D}}$ &=& 
           $ \pm  \rho_{(A}\sigma_{B)}\bar{\sigma}_{(\dot{C}}
                                     \bar{\rho}_{\dot{D})} $ & 5   \\[0.8ex]
F.   &    $ |(1,0)|^2   $    & $T_{AB\dot{C}\dot{D}}$ & =& 
           $ \pm  \rho_{A}\rho_{B} \bar{\rho}_{\dot{C}}
                                     \bar{\rho}_{\dot{D}} $   & 3 \\[0.8ex]
G.   &    $(0) $             &  $T_{AB\dot{C}\dot{D}}$ &=& $0$  & 0
\end{tabular}

\bigskip

Obviously this table shows that some cases can be obtained by 
specification of more general ones. This means for instance that 
the classes B1 and B2 can be obtained from 
A; or class F follows either from E or from B2. 

A further classification can be obtained by considering the 
eigenspaces of the tensor. For the purpose of this paper 
this refinement is not important. Details can be found in
\cite{PR,LS}.

\medskip

For example the energy-momentum tensor of a massless real scalar 
field $\phi$ is $T_{AB\dot{C}\dot{D}} = (\nabla_{A\dot{C}} \phi) 
 (\nabla_{B\dot{D}} \phi)$ and therefore in class C; in the 
massless spinor case we have $T_{AB\dot{C}\dot{D}} = \psi_A 
\nabla_{B\dot{C}} \bar{\psi}_{\dot{D}}$ in class D. 

The use of this classification to formulate matter actions in the 
sense of section 2 enables to discuss in full generality all 
possible matter sources for gravity and not only the ones 
treated in \cite{CDJ}. 

\bigskip
\bigskip

\section{Description of the constraints}

\bigskip

Since our aim is to formulate matter actions using the 2-form 
$\Sigma^{AB}$ as basic variable the entities of 
the Lagrangian (fields, derivatives, etc.) have to be written 
in the language of differential forms. The differential 
form may contain more physical degrees of freedom than one 
actually wants to describe. In this section we will 
present constraints that project out the undesired 
degrees of freedom. It is one of the ideas of this 
work to show how far one can come using only the 
tensor structure. Therefore the explicit dependence 
of the projection valued constraints on the 
physical fields needs not to be considered.

\medskip

For example a 1-form valued spinor field $\rho_A$ contains 
a spin 3/2 and a spin 1/2 part:
\beq
\rho_A = \rho_{AM\dot{M}} \theta^{M\dot{M}} = 
         \left(  \rho_{(AM)\dot{M}} + \epsilon_{AM} \tilde{\rho}_{\dot{M}}
         \right) \theta^{M\dot{M}}, \qquad 
 \rho_A \in (1 , 1/2) \oplus ( 0, 1/ 2 ).
\label{decomposition}
\eeq
We denote by ({\it i}, {\it j}) in the usual way the
finite-dimensional representations corresponding to $SL(2,{\bf C})$.

The undesired degrees of freedom have to be projected out by the use 
of constraints. The choice of the constraints should still 
allow a formulation of the action of the form (\ref{action3}). 
The sufficient cases for the present purposes are discussed. 
In what follows the object $\tau$ denotes the Lagrange multiplier.

\medskip

A 2-form $\rho^A$ contains components in the representations 
$({3 \over 2}, 0)$, $({1 \over 2}, 0 )$, and $ ( { 1\over 2}, 1)$. 
The following constraint can be considered:
\beq
\tau_{(AB)C} \Sigma^{AB} \rho^C ; \qquad \tau_{(AB)C}: \quad {\rm 0-form}.
\label{constraint1}
\eeq
By direct inspection and using \cite{PLEB} the orthogonality relation 
$\Sigma^{AB} \wedge \tilde{\Sigma}^{\dot{C}\dot{D}}=0$ 
it follows that the index symmetries of $ \tau_{(AB)C}$ decide 
which components can be projected out. 
\beq
\beqcol 
\tau_{(AB)C} = \tau_{(ABC)}   & \quad {\rm eliminates}
\quad & (3/2, 0)  \\
\tau_{(AB)C} = \tau_{[(AB)C]} & \quad {\rm eliminates}
\quad & (1/2, 0)  
\eeqcol
\label{constraint2}
\eeq
There is no constraint of the form (\ref{constraint1}) which 
can eliminate the $ ( { 1\over 2}, 1)$ component. This result is in
contrast to the supergravity case considered in \cite{CDJ}. There 
the first constraint of (\ref{constraint2}) has been taken. This
leaves besides of the desired Rarita-Schwinger field a 
Weyl-spinor component in the action. 

\medskip

If $\rho^A$ is taken as in (\ref{decomposition}) a good constraint 
is again of the form $ \tau_{(AB)C} \Sigma^{AB} \rho^C $. 
Here $ \tau_{(AB)C} $ has to be a 1-form. Like 
in (\ref{constraint2}) one gets:
\beq
\beqcol 
\tau_{(AB)C} = \tau_{(ABC)}  & \quad {\rm eliminates}
\quad & (1, 1/2)  \\
\tau_{(AB)C} = \tau_{[(AB)C]}  & \quad {\rm eliminates}
\quad & (0,1/2)  
\eeqcol
\label{constraint3}
\eeq 
In this case the advantage is that each component of 
$\rho^A$ can be projected out. 

\medskip

Given a 2-form $\rho^{(AB)} \in (2,0) \oplus (1,1) \oplus (1,0) 
\oplus (0,0)$ the constraint $\tau \Sigma_{AB} \rho^{AB} $ 
eliminates uniquely the $(0,0)$ component whereas 
$\tau_{(AB)(CD)}\Sigma^{AB} \rho^{CD} $ projects out exactly one 
of the components $(2,0)$, $(1,0)$, or $(0,0)$ obviously depending 
on the symmetries of $\tau$ among the pairs $(AB)$ and $(CD)$. 

\medskip

The last case important for this paper is a 1-form 
$ \rho^{AB} \in (3/2,1/2) \oplus (1/2,1/2)$. Here 
the condition $\tau \Sigma_{AB} \rho^{AB} $ allows 
for getting rid of the $(1/2,1/2)$ part. Another possible 
constraint is $\tau_{(AB)(CD)}\Sigma^{AB} \rho^{CD} $ with 
$\tau$ being a 1-form. Here we get:
\beq
\beqcol 
\tau_{(AB)(CD)} = \tau_{(ABCD)}   & \quad {\rm eliminates}
\quad & (3/2, 1/2)  \\
\tau_{(AB)(CD)} \ne \tau_{(ABCD)}    & \quad {\rm eliminates}
\quad & (1/2,1/2)  
\eeqcol
\label{constraint4}
\eeq

\bigskip
\bigskip

\section{Matter action in the self-dual 2-form formulation}

\bigskip

Now the question of how to construct matter actions in the 
scheme of section 2 is addressed. Notice that in this paper 
only trace-free energy-momentum tensors are being considered. Thus 
mass terms etc. are not treated but this extension 
is straightforward, see \cite{CDJ,PEL}.

We consider as an ansatz an action of the form: 
\beq 
S_{ansatz} = \int \Sigma^{AB} \Omega_{AB} 
\label{e1}
\eeq

$\Omega_{AB}$ should contain the matter degrees of freedom. 
However, its explicit dependence on physical fields is 
not needed for the discussion. 
Due to its tensor structure $\Omega_{AB}$ contains the 
$SL(2,{\bf{C}})$ irreducible components 
(2,0), (1,0), (0,0), and (1,1). Only the last 
term can be related to the traceless part of the 
energy-momentum tensor. 

The reducible energy-momentum tensors of section 3 can be obtained if 
we have:
\beq
\Omega_{AB} = \rho_A\sigma_B, \quad \rho_A F \sigma_B, \quad 
              F\Gamma_{AB}, \quad L_{AC} K^{C}_{\ \ B} .
\label{e2}
\eeq

Since $\Omega_{AB}$ is a 2-form the differentials can be assigned 
in various ways to its constituents in (\ref{e2}). One first 
decomposes the forms into their irreducible parts according to 
(\ref{decomposition}). 
Then the result is expanded with respect to the 
fundamental 2-forms using (\ref{expan1}). 
In general one then gets from (\ref{e1}) the expression:
\beq
\beqcol
S_{ansatz} = & \int& \Sigma^{AB} \Omega_{(AB)(\dot{C}\dot{D})} 
\tilde{\Sigma}^{\dot{A}\dot{B}}                      \\
  &+&  \sum \left(    ( 
       {\Omega}_{(ABCD)}  + 
            \epsilon_{AC} \Omega_{X (B \ D)}^{\ \ \ \ X}       
        + \epsilon_{AD} \Omega_{X (B C)}^{\ \ \ \ \ \ \ \ X} ) 
             \Sigma^{AB}\Sigma^{CD}    + 
      {\Omega}_{AB}^{AB} \Sigma_{RS}\Sigma^{RS}   \right)
\eeqcol
\label{expan2}
\eeq
The sum refers to the fact that the constituents of $\Omega$ 
may split into components which contribute 
to (\ref{expan2}) seperately. As above the same symbols are used 
for the forms itself and its irreducible components. 

In general one has to be careful about the completely symmetric 
terms. If $\Omega_{AB}$ can be split, the following 
subtlety in the (2,0) component might occur.
\beq
\Omega_{(AB)(MN)} = X_{(MN)C} \epsilon^{CD} Y_{(DAB)} 
                  = {}_a X_{(MNC)}\epsilon^{CD} Y_{(DAB)} 
                    + {}_bX_M Y_{(NAB)},
\label{irred}
\eeq
where ${}_aX$ and ${}_bX$ denote the irreducible 
components of the constituent $X$. The last term is completely 
symmetric and yields an 
additional term in (\ref{expan2}) which cannot be seen 
directly by the symmetries of $\Omega_{AB}$ itself.

\medskip

Within the framework of section 2 the Einstein equations 
are recovered by variation of the action with respect to 
$\Sigma^{AB}$. Therefore only the first term in (\ref{expan2}) 
should contribute since it contains the energy-momentum 
tensor of matter and in the sense of (\ref{einstein}) it is 
the correct source for gravity. Hence when really formulating an action 
for matter fields in the formalism of \cite{CDJ} the 
remaining terms in (\ref{expan2}) have to be subtracted 
consistently:
\beq  
\beqcol
S_{matter} = &\int& \Sigma^{AB} \Omega_{AB}        \\  
 &-& {1\over2} \sum \left(  ( {\Omega}_{(ABCD)} + 
            \epsilon_{AC} \Omega_{X (B \ D)}^{\ \ \ \ \ X}
          +  \epsilon_{AD} \Omega_{X (B C)}^{\ \ \ \ \ \ \ X} ) 
\Sigma^{AB}\Sigma^{CD}
    +  {\Omega}_{AB}^{AB} \Sigma_{RS}\Sigma^{RS}      \right)
\eeqcol
\label{expan3}
\eeq 
This form of the matter action allows for implementing 
the different irreducibility classes of the energy-momentum 
tensor.

After having established the equations for $\Sigma^{AB}$ 
in the gravity-matter system the (1,0) component of $\Omega_{AB}$ 
drops out and (\ref{expan3}) takes a form comparable to 
(\ref{action3}) and  (\ref{genaction1}).

\bigskip

Now it can be shown how the different irreduciblity classes of 
the enery-momentum tensor can be brought in an action of the form 
(\ref{expan3}).

\medskip

There are two possibilities for the classes B1, B2, and C 
which may e.g. describe the massless scalar field. 

The first 
is given by $\Omega=F\Gamma_{AB}$ with both $F$ and $\Gamma_{AB}$ 
being 1-forms. According to (\ref{constraint4}) the action 
is given by:
\beq
S = \int \Sigma^{AB} F \Gamma_{AB}  
    + {1\over 4} F_{C \dot{C}} \Gamma^{C \dot{C}} 
    \Sigma_{AB}\Sigma^{AB} + \tau_{(ABCD)}\Sigma^{AB}\Gamma^{CD}
\label{vect1}
\eeq
The second possiblity is provided by the decomposition 
into two 1-forms: $\Omega_{AB} = L_{AC} K^{C}_{B}
$. In this case two constraints of kind (\ref{constraint4}) 
are required. These may coincide depending on the actual class.
\beq
S = \int \Sigma^{AB} L_{AC} K^{C}_{B} 
  -  {1\over 4}  L_{C \dot{C}} K^{C \dot{C}} \Sigma_{AB}\Sigma^{AB} 
   +\Sigma^{AB} \left(  \tau_{(ABCD)}L^{CD} + 
                        \upsilon_{(ABCD)}K^{CD} \right) 
\label{vect2}
\eeq

Since the actions for the classes B1, B2, and C do contain 
trace parts a pure connection formulation \cite{CDJ} for 
the gravity-matter system becomes difficult even if one 
does not include explicitly traces of the energy-momentum tensor (see below).

\bigskip

Next the class D is considered. The energy-momentum tensor of 
a massless spin $1/2$ particle is within this class. However, 
the occurence of torsion will not be treated. 

This class can be described by $\Omega_{AB} = \rho_A\sigma_B$ 
with either $\rho_A$ a 2-form and $\sigma_B$ a 0-form 
or both being 1-forms. In both cases two constraints for 
projecting onto the desired physical degrees of freedom 
are necessary. The result for the first possiblity is:
\beq
S = \int \Sigma^{AB} \rho_{A} \sigma_{B} 
           +  \Sigma^{AB} \left( \tau_{(ABC)}\rho^{C} + 
                        \upsilon_{[(AB)C]}\sigma^{C} \right) 
\label{spinor1} 
\eeq
Notice that no additional subtractions have to be performed. 
One gets a similar result for the above mentioned second 
possibility. 

A third description for the class D similar to the one in 
\cite{CDJ} is possible. One takes 
$\Omega_{AB} = \rho_A F \sigma_B$. Here $\rho$ and $F$ are 
1-forms. The resulting action is:
\beq
S = \int \Sigma^{AB} \rho_{A}F \sigma_{B} 
           + {1\over 4} \rho_{\dot{C}} F_{D}^{\ \ \dot{C}}\sigma^D 
\Sigma^{AB}\Sigma_{AB}
+  \tau_{(ABC)}\Sigma^{AB}\rho^C  
\label{spinor2}
\eeq

It is interesting to note that while (\ref{spinor1}) permits 
straightforwardly a pure connection formulation the 
action (\ref{spinor2}) causes problems.

\bigskip

For the classes E and F the description has already been 
given in \cite{CDJ}. The Yang-Mills field is within these classes. 
One has to choose $\Omega_{AB}=\rho_A \sigma_B F$. $F$ has to be a 2-form. 
According to the decomposition given in (\ref{expan2}) one gets 
$\Omega_{AB} = \rho_A \sigma_B ( F_{(CD)}  \Sigma^{CD} 
               + \tilde{F}_{(\dot{C}\dot{D})} 
              \tilde{ \Sigma}^{\dot{C}\dot{D}}  )$. 
The two components of $F$ correspond to its 
$(1,0)$ and $(0,1)$ parts respectively. 

Since only the second term will give a correct contribution to 
the energy-momentum tensor the action in this case is
\beq
S =  \int \Sigma^{AB}\rho_A \sigma_B F - {1\over 2} 
         \rho_A \sigma_B F_{(CD)} \Sigma^{AB} \Sigma^{CD}
\label{YM}
\eeq
The hermiticity requirement for the energy momentum 
tensor relates $F_{(CD)}$ to $ \rho_A \sigma_B$. A recent 
treatment of the Einstein-Yang-Mills system \cite{ROB} is 
equivalent to the one given here or in \cite{CDJ}. However, 
it can be read off (\ref{YM}) that after establishing 
the equations of motion for $\Sigma^{AB}$ the second 
terms becomes a trace. Therefore a pure 
connection formulation for classes E and F is difficult.

\medskip

Formally one could obtain for all the above mentioned 
matter actions a pure spin-connection formulation. This 
could be achieved by simply preforming the following 
replacements in (\ref{revact1}):

\beq
\beqcol
R_{AB} & \longrightarrow& R_{AB} + {\rm matter \ \ terms \ \ linear \ \ 
in} \ \  \Sigma \\
X & \longrightarrow & X   + {\rm matter \ \ terms \ \ quadratic \ \ 
in} \ \ \Sigma 
\eeqcol
\eeq

After having done these substitutions one could in 
principle perform the procedure at the end 
of section2 to obtain the pure spin-connection formulation. 
But in doing this one faces in addition to 
the pathological points in that procedure the serious problem 
that the matter 
terms are of local character. It is therefore 
not guaranteed globally that the operations 
involving $X$ in section 2 are possible everywhere 
in space time !

\medskip

An almost unique description for the reducible classes of 
the energy-momentum tensor has been obtained. The 
irreducible ones involve fields of at least spin 3/2. 
For physical reasons these higher spin systems coupled 
to gravity are problematic although formal actions can 
easily be formulated within the presented scheme. 

Only a formulation for the spin 3/2 field is 
given as a final example. One combines a 
2-form $\rho_C$ and a 0-form $\sigma_{(CAB)}$ to 
$\Omega_{AB} = \rho_C \sigma_{\ \ AB}^C $. The object 
$\sigma_{(CAB)}$ then consists of the contracted product of 
a derivative and one component of the spin 3/2 field.

In (\ref{constraint2}) 
it was shown that two constraints are needed for $\rho_A$ 
in order to describe a Rarita-Schwinger field. The action 
in contrast to \cite{CDJ} is:
\beq
S = \int \Sigma^{AB} \rho_C \sigma_{\ \ AB}^C + 
    \Sigma^{AB} ( \tau_{(ABC)} + \upsilon_{[(AB)C]} ) \rho^C
\label{rarita}
\eeq
 
\bigskip
\bigskip

\section{Remarks on scalar- and spinor-field actions}

\bigskip

It is necessary to make a remark about the actions 
for a scalar field and a spin $1/2$ field. 
This is because in comparing the spinor-field actions 
of \cite{CDJ} and \cite{PEL} one might wonder why they 
differ by constraint terms. It will be argued that 
the use of actions involving constraints like in section 5 is 
not really necessary. 

One can rewrite the standard action of a scalar field coupled to 
gravity using \cite{URB,CDJ}:
\beq
\sqrt{|g|}g_{mn} = {1\over 3} \epsilon^{abcd} 
          \Sigma^{AB}_{ma}\Sigma_{bcB}^{\ \ \ \ C} 
          \Sigma_{dnCA}
\label{urb}
\eeq
Using (\ref{vect2}) one can check that

\beq
S = \int \Sigma_{A}^{\ \ B}  L_{BC} K^{CA} - {1\over 2}
    \Sigma_{RS}\Sigma^{RS} \tilde{L}_{A\dot{B}} \tilde{K}^{A\dot{B}}
\label{actscal}
\eeq
reproduces with the help of (\ref{urb}) the standard 
scalar-field action in curved space. In the sense of 
(\ref{decomposition}) we denote by $ \tilde{L}_{A\dot{B}} $ and 
$\tilde{K}_{A\dot{B}}$ the vectorial (1/2,1/2) components 
of the 1-forms $L_{BC}$ and $K^{CA}$ respectively.

(\ref{vect2}). The constraints needed in (\ref{vect2}) 
drop out because the use of formula (\ref{urb}) 
restricts the 1-forms $K$ and $L$:
\beq
\beqcol
K^{AB} &=& \Sigma^{AB}_{mn}\partial^m \phi {\rm d}x^n 
         = \tilde{K}^{(A}_{\ \ \dot{A}}\theta^{B)\dot{A}}      \\
L^{AB} &=& \Sigma^{AB}_{mn}\partial^m \psi {\rm d}x^n   
          = \tilde{L}^{(A}_{\ \ \dot{A}}\theta^{B)\dot{A}} 
\eeqcol
\label{actscalar2}
\eeq

$\phi$ and $\psi$ are scalar fields which may coincide, i.e. 
$\tilde{K}_{A\dot{A}} = \nabla_{A\dot{A}}\phi$ and 
$\tilde{L}_{B\dot{B}} = \nabla_{B\dot{B}}\psi$

\medskip

If one now applies the obvious restriction from 
class B1 to class D of the energy-momentum tensor in the action 
it is clear that also the action for the spin $1/2$ field can be 
formulated without using specific constraints as it 
has been done in \cite{PEL}. The 
trace-term plays no role in this case due to the equations of 
motion for the spinor field.

\bigskip
\bigskip

{\bf Acknowledgement}

Discussions with Professors H. Kodama and R. Sasaki are 
gratefully acknowledged. Thanks also to an unknown referee 
whose comments helped to improve the paper.

\end{document}